# Electrostatic carrier doping of GdTiO$_3$/SrTiO$_3$ interfaces

Pouya Moetakef[1], Tyler A. Cain[1], Daniel G. Ouellette[2], Jack Y. Zhang[1], Dmitri O. Klenov[3], Anderson Janotti[1], Chris G. Van de Walle[1], Siddharth Rajan[4], S. James Allen[2], and Susanne Stemmer[1]

[1] Materials Department, University of California, Santa Barbara, California, 93106-5050, USA

[2] Department of Physics, University of California, Santa Barbara, California, 93106-9530, USA

[3] FEI, Achtseweg Noord 5, 5651 GG Eindhoven, Netherlands

[4] Department of Electrical and Computer Engineering, The Ohio State University, Columbus, OH 43210 USA

**Abstract**

Heterostructures and superlattices consisting of a prototype Mott insulator, GdTiO$_3$, and the band insulator SrTiO$_3$ are grown by molecular beam epitaxy and show intrinsic electronic reconstruction, approximately ½ electron per surface unit cell at each GdTiO$_3$/SrTiO$_3$ interface. The sheet carrier densities in all structures containing more than one unit cell of SrTiO$_3$ are independent of layer thicknesses and growth sequences, indicating that the mobile carriers are in a high concentration, two-dimensional electron gas bound to the interface. These carrier densities closely meet the electrostatic requirements for compensating the fixed charge at these polar interfaces. Based on the experimental results, insights into interfacial band alignments, charge distribution and the influence of different electrostatic boundary conditions are obtained.

Two-dimensional electron gases (2DEGs) at interfaces between Mott insulators and band insulators have attracted significant attention because of unique properties, such as strong electron correlations, superconductivity or magnetism [1-7]. Furthermore, interfaces between the band insulator SrTiO$_3$ and the rare earth titanates ($R$TiO$_3$, where $R$ is a trivalent rare earth ion), which are Mott insulators, exhibit a fixed polar charge. In particular, $R^{3+}$O$^{2-}$ and Ti$^{3+}$O$_2^{4-}$ layers alternate along the (001) surface normal of $R$TiO$_3$ [8], carrying formal +1 and -1 charges, respectively, which causes a diverging electrostatic surface energy due to the non-zero dipole moment on the $R$O-TiO$_2$ units. At the interface, these transition to a sequence of neutral layers, Sr$^{2+}$O$^{2-}$ and Ti$^{4+}$O$_2^{4-}$, of non-polar (001) SrTiO$_3$. The fixed interfacial charge can be compensated by a 2DEG, residing in the bands of the Mott and/or band insulator and bound to the interface by the fixed charge [9,10]. In the absence of any other charge compensation, defects, interfacial mixing, roughness and nonstoichiometry [11-13], the interface is expected to form an extremely high-density 2DEG on the order of $3\times10^{14}$ cm$^{-2}$, as given by $e/2S$, where $S$ is the surface unit cell area and $e$ the elementary charge. The nature and spatial distribution of charge carriers are of paramount importance for the properties of these heterostructures.

To date, attention has focused on LaAlO$_3$/SrTiO$_3$ and LaTiO$_3$/SrTiO$_3$ interfaces grown by pulsed laser deposition [1,14-16]. Results from electrical transport measurements vary significantly; in particular, LaAlO$_3$/SrTiO$_3$ interfaces show carrier densities that are an order of magnitude less than predicted from intrinsic electronic reconstruction [17-19]. Compensating mobile electrons are easily accessible for $R$TiO$_3$/SrTiO$_3$ structures, as can be visualized by considering the atomically sharp interface as a 50:50 mixture of $R$TiO$_3$ and SrTiO$_3$, which has the required free electron density [20]. Transport and optical measurements of LaTiO$_3$/SrTiO$_3$ interfaces reveal densities close to those expected for electronic reconstruction [16,21], but interpretation is complicated by conduction by non-interfacial carriers from both substrates and films [15,16]; LaTiO$_3$ films reported in the literature are often metallic [22]. This Letter reports on transport measurements of the 2DEGs at GdTiO$_3$/SrTiO$_3$ interfaces (electronically analogous to LaTiO$_3$/SrTiO$_3$ [23,24]) grown by molecular beam epitaxy (MBE) that exhibit an interfacial density with values that are essentially those predicted by electronic reconstruction. Theoretical band offsets between the conduction bands of SrTiO$_3$ and GdTiO$_3$ are used to model the spatial extent of the 2DEG.

The vastly different oxygen pressures required to obtain insulating SrTiO$_3$ and GdTiO$_3$ layers present an experimental challenge: high oxygen pressure is needed for insulating SrTiO$_3$, while GdTiO$_3$ films need to be grown under low oxygen pressure conditions to avoid metallic conductivity or formation of pyrochlore [25,26]. We use (001) surfaces of (LaAlO$_3$)$_{0.3}$(Sr$_2$AlTaO$_6$)$_{0.7}$ (LSAT) as substrates to avoid substrate conduction. All layers and superlattices were grown by MBE. SrTiO$_3$ was grown by co-deposition [27], whereas GdTiO$_3$ was grown by shuttered growth, supplying alternating monolayer doses of Gd and Ti tetra isopropoxide (TTIP), which supplied both Ti and oxygen. No additional oxygen was supplied [28]. For GdTiO$_3$ on SrTiO$_3$, growths were started and terminated with a TiO$_2$ layer. All layers and superlattices were coherently strained to the LSAT [29]. Aberration-corrected scanning transmission electron microscopy (FEI Titan G2 ChemiSTEM) was used to characterize the atomic structure of GdTiO$_3$/SrTiO$_3$ interfaces. Longitudinal and Hall resistivity were measured in Van der Pauw geometry using a Physical Properties Measurement System (Quantum Design

PPMS). Ohmic contacts were 300 nm Au/20 nm Ni/40 nm Al for SrTiO$_3$ top layers and 300 nm Au/50 nm Ti for GdTiO$_3$ top layers. The top layer was Au for wire bonding with an Au wire.

The sheet resistances of GdTiO$_3$ grown directly on LSAT and of GdTiO$_3$ grown on SrTiO$_3$ buffer layers with different thicknesses are shown in Fig. 1(a). The GdTiO$_3$ film on LSAT with no SrTiO$_3$ buffer layer is insulating. While too resistive for meaningful Hall measurements, the Seebeck coefficient was measured and is positive (p-type), as found for stoichiometric GdTiO$_3$ [23]. All bilayers are *n*-type and metallic if the SrTiO$_3$ thickness exceeded one unit cell (0.4 nm). Even the bilayer with one unit cell SrTiO$_3$ already exhibits a remarkable drop in resistance. The localized behavior for this sample is expected as the sheet resistance exceeds the critical Mott value (~10 kΩ/□). The sheet resistance should decrease with increasing SrTiO$_3$ thickness if the conductivity is due to the oxygen deficient SrTiO$_3$. The constant sheet resistance for SrTiO$_3$ layers thicker than 20 nm indicates that it arises from a space charge layer of constant thickness and carrier density at the interface. The Hall resistance as a function of magnetic field *B* was linear and *n*-type down to the lowest temperatures [29], in contrast to LaTiO$_3$/SrTiO$_3$ [15,16]. All of the electrons contributing to the Hall resistance satisfy $\mu B \ll 1$. Although more than one subband with different mobility may be occupied, the Hall coefficient ($R_H$) is converted to an effective sheet density by $n_S = 1/eR_H$, where *e* is the elementary charge. Figure 1(b) shows that $n_S$ is constant, ~ $3.5 \times 10^{14}$ cm$^{-2}$, for all bilayers, even for extremely thin SrTiO$_3$. Thus there is little trapping at the LSAT/SrTiO$_3$ interface, at least on a scale of ~ $3 \times 10^{14}$ cm$^{-2}$. A similar result is obtained when the GdTiO$_3$ thickness is varied. The mobility increases with SrTiO$_3$ thickness [Fig. 1(b)].

Thus the decrease in sheet resistance for SrTiO$_3$ layers between 0.4 nm and 20 nm is due to an increase in mobility, not a change in sheet carrier concentration. The results are consistent with an interfacial mobile space charge layer of constant thickness with a sheet charge density of ~ $3\times10^{14}$ cm$^{-2}$. This carrier density closely corresponds to the ½ electron per surface unit cell required to compensate for the polar discontinuity at the interface.

To further confirm the results, multilayer samples were investigated. Figure 2(a) shows the sheet carrier density for (SrTiO$_3$/GdTiO$_3$/SrTiO$_3$)$_x$ superlattices on LSAT as a function of the number of repeats $x$, each containing two GdTiO$_3$/SrTiO$_3$ interfaces. If each repeat contributes the same sheet carrier density as the $x$ = 1 trilayer, then the sheet density should scale as shown by the dashed line. The experimental results closely follow the dashed line, independent of GdTiO$_3$ or SrTiO$_3$ thicknesses. The total carrier sheet density is not proportional to the total Gd in the superlattice nor is it proportional to the Gd concentration. It is proportional to the number of interfaces. The constant sheet carrier density per interface excludes interfacial intermixing as the source of the charge carriers, because the charge carrier density in this case should strongly dependent on the precise interfacial composition [20]. Figure 2(b) shows a high-angle annular dark-field scanning transmission electron microscopy (HAADF/STEM) image of the sample with $x$ = 20. Thicker sample regions appear to show intermixing of about one monolayer, but observation of thinner regions [29] shows that the interface contains short steps, which overlap along the beam direction. Thus the interfaces are *locally* atomically abrupt. Comparison of sheet carrier densities of the $x$ = 1 multilayer (two GdTiO$_3$/SrTiO$_3$ interfaces) with that of the GdTiO$_3$/SrTiO$_3$/LSAT structures (one interface) of Fig. 1(b), shows that the sheet carrier concentration of the multilayer is slightly less than twice that of a single interface. This is likely due to different electrostatic boundary conditions for GdTiO$_3$/SrTiO$_3$ and SrTiO$_3$/GdTiO$_3$ interfaces.

The interfacial space charge can be understood by noting that the fixed polar charge at the interface must be neutralized by negative space charge, mobile or fixed, and dictated by the available quantum states in the presence of the selfconsistent electrostatic fields/potentials. The SrTiO$_3$ is *n*-type (oxygen deficient) and GdTiO$_3$ is *p*-type. The fixed polar charge can be neutralized by an accumulation layer in the SrTiO$_3$, a hole depletion layer (negatively charged acceptors) in the GdTiO$_3$ and an inversion layer in the GdTiO$_3$. The mobile charge is close to that required to compensate the fixed polar charge at the interface: thus hole depletion in the GdTiO$_3$ is not sufficient to siphon off significant numbers of electrons from the mobile space charge. The interface may share the mobile charge between the SrTiO$_3$ and the GdTiO$_3$. The relatively strong temperature dependence of the electron mobility [Fig. 1(b)] and the absence of an anomalous Hall effect, potentially caused by the ferrimagnetism in the GdTiO$_3$, indicate that mobile charge is largely found on the SrTiO$_3$ side and that the conduction band alignment favors SrTiO$_3$ accumulation. Because SrTiO$_3$ has the larger band gap, the band line-up must be of type II (staggered). First principle calculations confirm this [29]. The mobile charge distribution and band bending are modeled using a self consistent Poisson-Schrödinger solver [30], as shown in Fig. 3, using band lineups from first-principle calculations [29], a fixed interface charge of $3.4 \times 10^{14}$ cm$^{-2}$ (modeled as a 0.2 nm layer with fully ionized dopants), an electron effective mass of 1 $m_0$, a 6-fold degeneracy of the conduction band, and dielectric constants of SrTiO$_3$ and GdTiO$_3$ of 300 and 30, respectively. The interfacial positive charge induces a high-density 2DEG. A deep quantum well is formed [Fig. 3(b)], but there is overflow of the electrons into the GdTiO$_3$. Despite the high effective mass/density of states, the high electron density drives the Fermi level above the SrTiO$_3$ conduction band minimum by approximately 0.7 eV, which is greater than the assumed conduction band offset. The GdTiO$_3$ conduction band is therefore near the Fermi level near the interface and the polar charges inverts the *p*-type GdTiO$_3$ ($N_A$ = $3 \times 10^{19}$ cm$^{-3}$), making it effectively *n*-type. From the simulations, the spatial extent of the quantum confined electron gas is ~ 3 nm. We note that the superlattice with only 4 nm SrTiO$_3$ is best described as a quantum well rather than two distinct interface space charge layers, yet the total electron density appears fixed by the polarization charge.

The model supports experimental observations, namely that the mobile space charge density at the GdTiO$_3$/SrTiO$_3$ interface is perturbed very little by the LSAT even for small separations and that the transport is dominated by one carrier type. The electrical transport measurements indicate that the different 2DEG regions in multi-layer structures are not isolated, which may have been expected since the GdTiO$_3$ layers are *p*-type. Most importantly, the very tight binding of the electrons to the interface should allow for exploration of quantum and strong correlation effects. Figure 3 is based on an effective mass model that assumes slowly varying envelope wave functions. The wavefunctions are derived from *d*-bands that are likely better described by tight binding Hamiltonians with rapid spatial variations, far from the approximations used in conventional semiconductor heterostructures. Appropriate models need to be developed, especially those that also include electron correlations.

P.M. and J.Y.Z. were supported through awards from the U.S. National Science Foundation (Grant No. DMR-1006640) and DOE Basic Energy Sciences (Grant No. DEFG02-02ER45994).  T.A.C. was supported through the Center for Energy Efficient Materials, an Energy Frontier Research Center funded by the DOE (Award Number DE-SC0001009).  S.S., S.J.A. and D.G.O. acknowledge support through a MURI program of the Army Research Office (Grant No. W911-NF-09-1-0398).  The work made use of the UCSB Nanofabrication Facility, a part of the NSF-funded NNIN network.

# References

[1]  A. Ohtomo, D. A. Muller, J. L. Grazul, and H. Y. Hwang, Nature **419**, 378 (2002).

[2]  S. Okamoto and A. J. Millis, Nature **428**, 630 (2004).

[3]  C. Yoshida, H. Tamura, A. Yoshida, Y. Kataoka, N. Fujimaki, and N. Yokoyama, Jpn. J. Appl. Phys. Part 1 **35**, 5691 (1996).

[4]  S. e. Smadici, P. Abbamonte, A. Bhattacharya, X. Zhai, B. Jiang, A. Rusydi, J. N. Eckstein, S. D. Bader, and J.-M. Zuo, Phys. Rev. Lett. **99**, 196404 (2007).

[5]  J. Garcia-Barriocanal, F. Y. Bruno, A. Rivera-Calzada, Z. Sefrioui, N. M. Nemes, M. Garcia-Hernandez, J. Rubio-Zuazo, G. R. Castro, M. Varela, S. J. Pennycook, Carlos Leon, and J. Santamaria, Adv. Mater. **22**, 627 (2010).

[6]  J. Garcia-Barriocanal, J. C. Cezar, F. Y. Bruno, P. Thakur, N. B. Brookes, C. Utfeld, A. Rivera-Calzada, S. R. Giblin, J. W. Taylor, J. A. Duffy, S. B. Dugdale, T. Nakamura, K. Kodama, C. Leon, S. Okamoto, and J. Santamaria, Nature Commun. **1**, 82 (2010).

[7]  Z. S. Popovic and S. Satpathy, Phys. Rev. Lett. **94**, 176805 (2005).

[8]  Pseudocubic notation is used here for the orthorhombic $RTiO_3$.

[9]  G. A. Baraff, J. A. Appelbaum, and D. R. Hamann, Phys. Rev. Lett. **38**, 237 (1977).

[10] W. A. Harrison, E. A. Kraut, J. R. Waldrop, and R. W. Grant, Phys. Rev. B **18**, 4402 (1978).

[11] S. A. Chambers, Surf. Sci. **605**, 1133 (2011).

[12] D. G. Schlom and J. Mannhart, Nat. Mater. **10**, 168 (2011).

[13] D. O. Klenov, D. G. Schlom, H. Li, and S. Stemmer, Jpn. J. Appl. Phys. Part 2 **44**, L617 (2005).

[14] J. Mannhart, D. H. A. Blank, H. Y. Hwang, A. J. Millis, and J. M. Triscone, MRS Bulletin **33**, 1027 (2008).

[15] R. Ohtsuka, M. Matvejeff, K. Nishio, R. Takahashi, and M. Lippmaa, Appl. Phys. Lett. **96**, 192111 (2010).

[16] J. S. Kim, S. S. A. Seo, M. F. Chisholm, R. K. Kremer, H. U. Habermeier, B. Keimer, and H. N. Lee, Phys. Rev. B **82**, 201407 (2010).

[17] M. Huijben, A. Brinkman, G. Koster, G. Rijnders, H. Hilgenkamp, and D. H. A. Blank, Adv. Mater. **21**, 1665 (2009).

[18] A. D. Caviglia, S. Gariglio, N. Reyren, D. Jaccard, T. Schneider, M. Gabay, S. Thiel, G. Hammerl, J. Mannhart, and J. M. Triscone, Nature **456**, 624 (2008).

[19] S. Thiel, G. Hammerl, A. Schmehl, C. W. Schneider, and J. Mannhart, Science **313**, 1942 (2006).

[20] Y. Tokura, Y. Taguchi, Y. Okada, Y. Fujishima, T. Arima, K. Kumagai, and Y. Iye, Phys. Rev. Lett. **70**, 2126 (1993).

[21] S. S. A. Seo, W. S. Choi, H. N. Lee, L. Yu, K. W. Kim, C. Bernhard, and T. W. Noh, Phys. Rev. Lett. **99**, 266801 (2007).

[22] A. Schmehl, F. Lichtenberg, H. Bielefeldt, J. Mannhart, and D. G. Schlom, Appl. Phys. Lett. **82**, 3077 (2003).

[23] H. D. Zhou and J. B. Goodenough, J. Phys.: Condens. Matter **17**, 7395 (2005).

[24] D. A. Crandles, T. Timusk, J. D. Garrett, and J. E. Greedan, Physica C **201**, 407 (1992).

[25]  A. Ohtomo, D. A. Muller, J. L. Grazul, and H. Y. Hwang, Appl. Phys. Lett. **80**, 3922 (2002).

[26]  K. S. Takahashi, M. Onoda, M. Kawasaki, N. Nagaosa, and Y. Tokura, Phys. Rev. Lett. **103**, 057204 (2009).

[27]  B. Jalan, R. Engel-Herbert, N. J. Wright, and S. Stemmer, J. Vac. Sci. Technol. A **27**, 461 (2009).

[28]  P. Moetakef, J. Y. Zhang, A. Kozhanov, B. Jalan, R. Seshadri, S. J. Allen, and S. Stemmer, Appl. Phys. Lett. **98**, 112110 (2011).

[29]  See supplemental material at [link to be inserted by publisher] for structural data, Hall resistance as a function of magnetic field and details of the first principle calculations of the $GdTiO_3$/$SrTiO_3$ band offsets.

[30]  M. Grundmann, BandEng program (unpublished).

**Figure Captions**

**Figure 1** (color online): (a) Sheet resistance as a function of temperature for GdTiO$_3$/SrTiO$_3$/LSAT structures with varying SrTiO$_3$ thicknesses, indicated by the labels. The GdTiO$_3$ film grown directly on LSAT is labeled "0 nm". (b) Sheet carrier density and mobility at room temperature and 2.5 K.

**Figure 2** (color online): (a) Room temperature sheet carrier concentrations of SrTiO$_3$/GdTiO$_3$/SrTiO$_3$ multilayers as a function of multilayer repeats ($x$). The dashed line indicates the expected sheet carrier concentration scaling with number of repeats as calculated from the $x = 1$ sample. (b) High-angle annular dark-field scanning transmission electron microscopy image of the $x = 20$ multilayer.

**Figure 3** (color online): Calculated (a) charge distribution and (b) band alignment for a SrTiO$_3$/GdTiO$_3$/SrTiO$_3$ heterostructure. The Fermi level is shown as a dotted line.

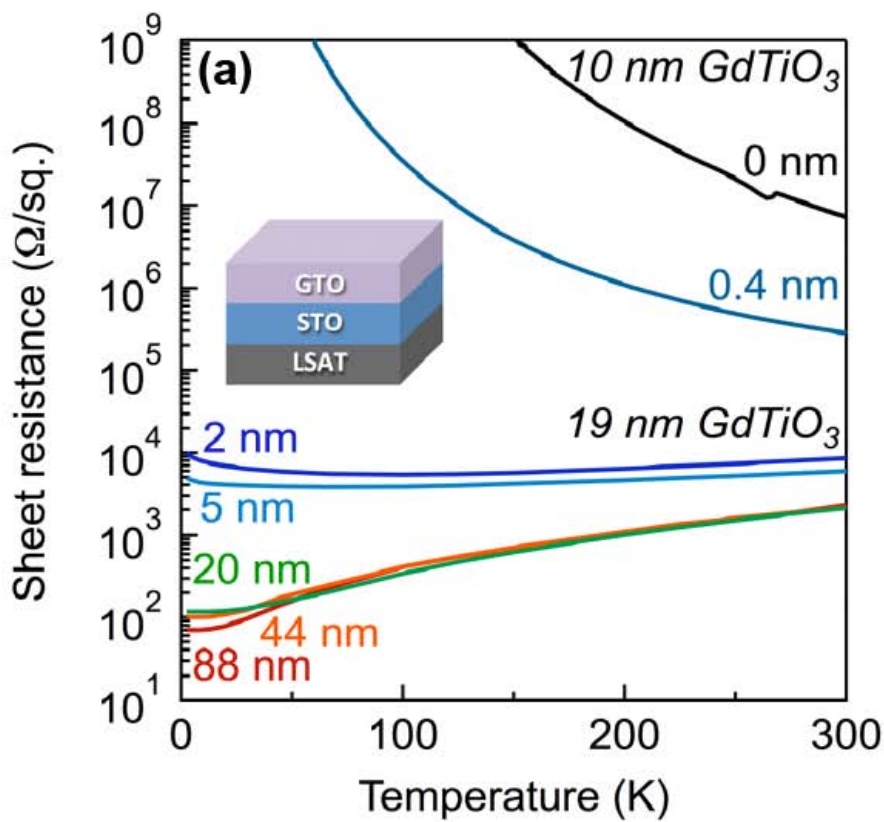
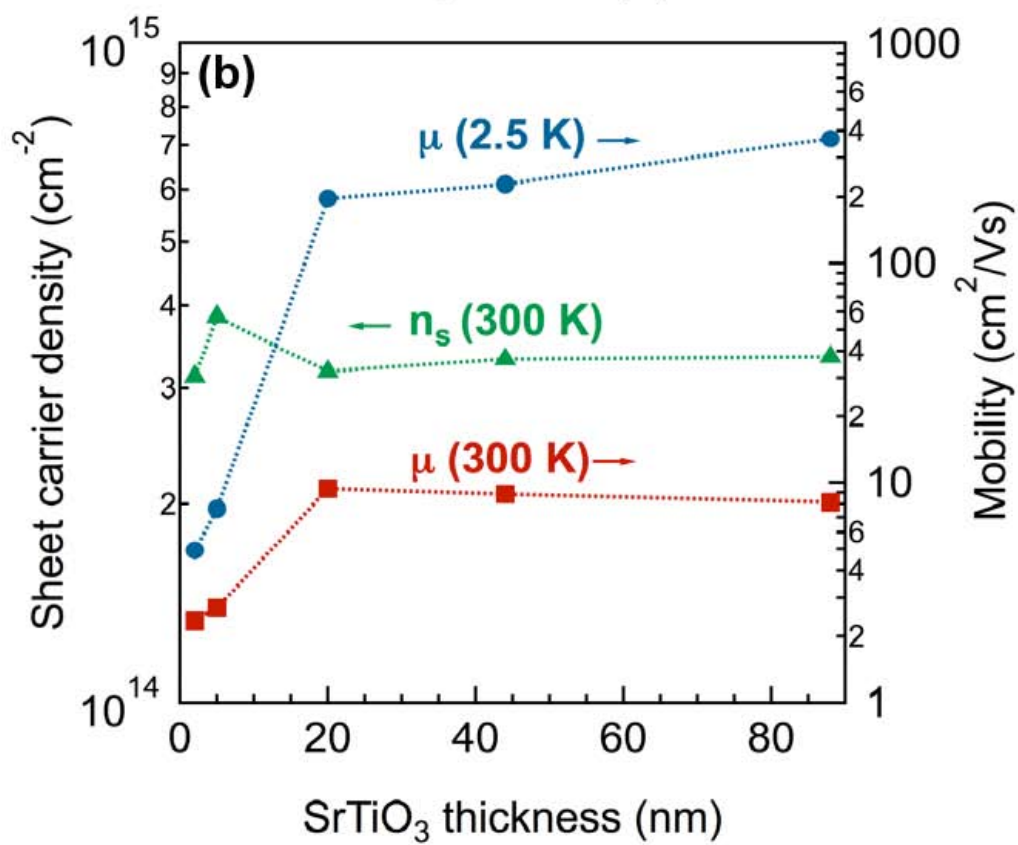

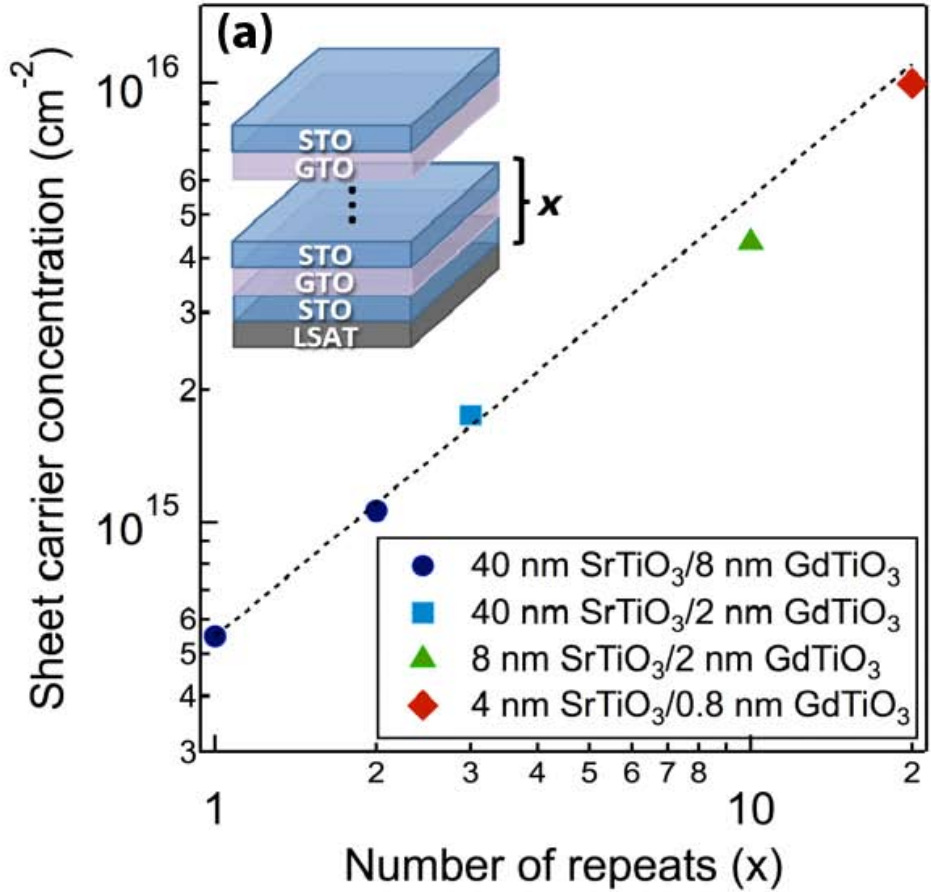
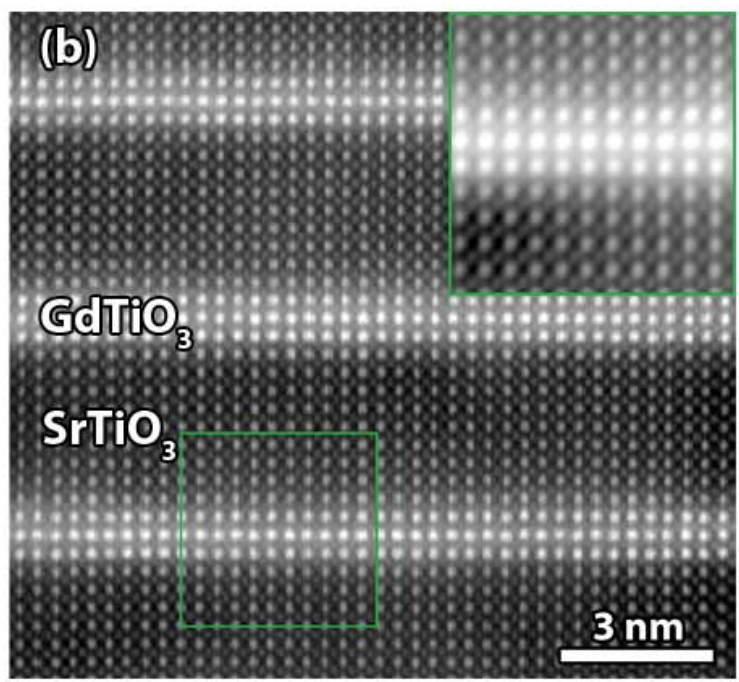

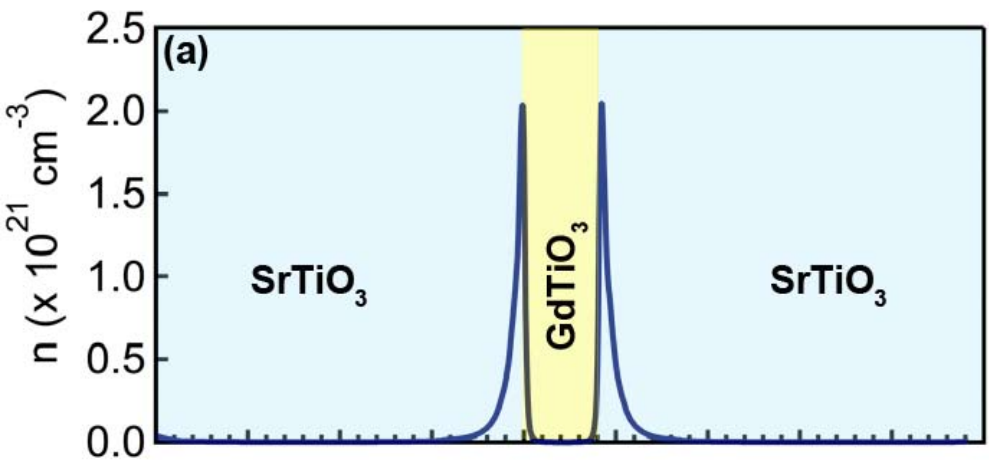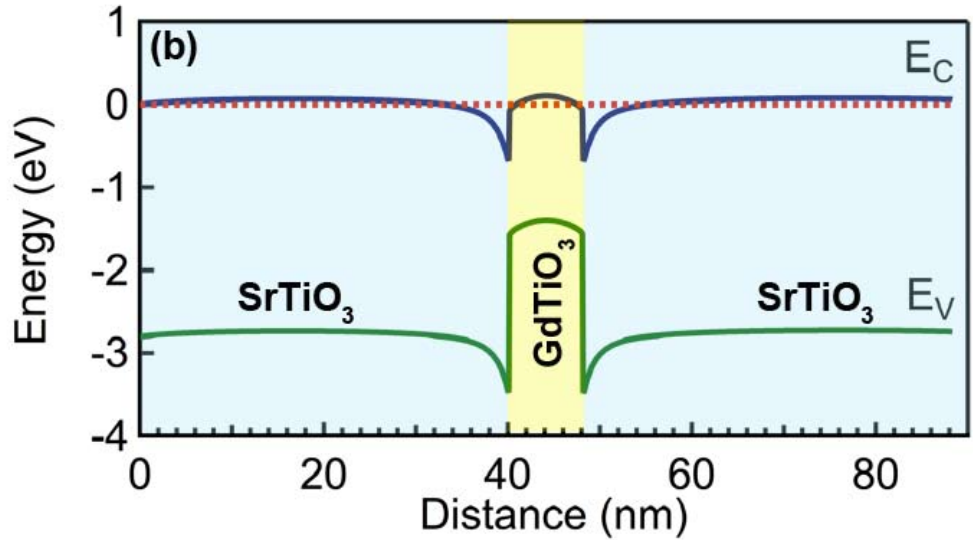

# Supplementary information: Electrostatic carrier doping of GdTiO$_3$/SrTiO$_3$ interfaces

Pouya Moetakef[1], Tyler Cain[1], Daniel G. Ouellette[2], Jack Y. Zhang[1], Dmitri O. Klenov[3], Anderson Janotti[1], Chris G. Van de Walle[1], Siddharth Rajan[4], S. James Allen[2], and Susanne Stemmer[1]

[1] Materials Department, University of California, Santa Barbara, California, 93106-5050, USA.
[2] Department of Physics, University of California, Santa Barbara, California, 93106-9530, USA.
[3] FEI, Achtseweg Noord 5, 5651 GG Eindhoven, Netherlands.
[4] Department of Electrical and Computer Engineering, The Ohio State University, Columbus, OH 43210, USA.

**Representative structural data from films and superlattices**

Representative structural data from GdTiO$_3$/SrTiO$_3$ samples are shown below. Figure S1 shows a x-ray reciprocal space map around the 103 LSAT substrate reflection of a (GdTiO$_3$/SrTiO$_3$)$_{10}$ superlattice with 2 nm GdTiO$_3$ and 8 nm SrTiO$_3$ layers. The substrate and superlattice peaks have the same in-plane lattice spacing, indicating that the superlattice is coherently strained to the substrate.

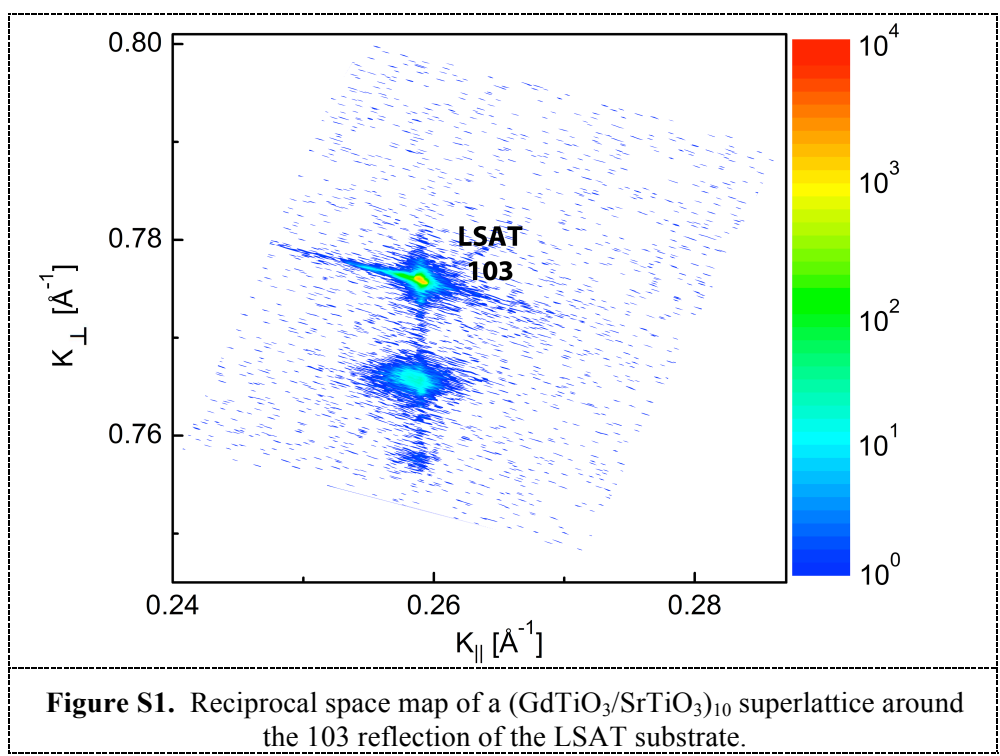

**Figure S1.** Reciprocal space map of a (GdTiO$_3$/SrTiO$_3$)$_{10}$ superlattice around the 103 reflection of the LSAT substrate.

Figure S2 shows a radial x-ray scan of the (GdTiO$_3$/SrTiO$_3$)$_{10}$ superlattice, showing the superlattice reflections.

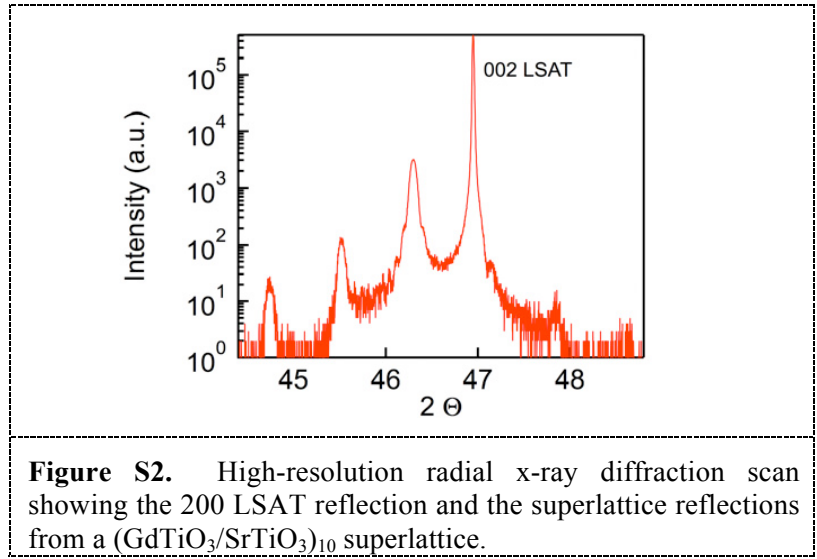

**Figure S2.** High-resolution radial x-ray diffraction scan showing the 200 LSAT reflection and the superlattice reflections from a $(GdTiO_3/SrTiO_3)_{10}$ superlattice.

Figure S3 shows reflection high-energy electron diffraction (RHEED) oscillations during the shuttered growth of a $GdTiO_3$ film on LSAT and RHEED patterns of the $GdTiO_3$ film surface after growth. Figure S4 show the corresponding RHEED data for $SrTiO_3$ film growth. The streaky RHEED patterns indicate atomically flat surfaces, consistent with the layer-by-layer growth mode. The 4× surface reconstruction (see arrows) observed along [110] of $SrTiO_3$ is a reliable indicator of cation stoichiometry [1].

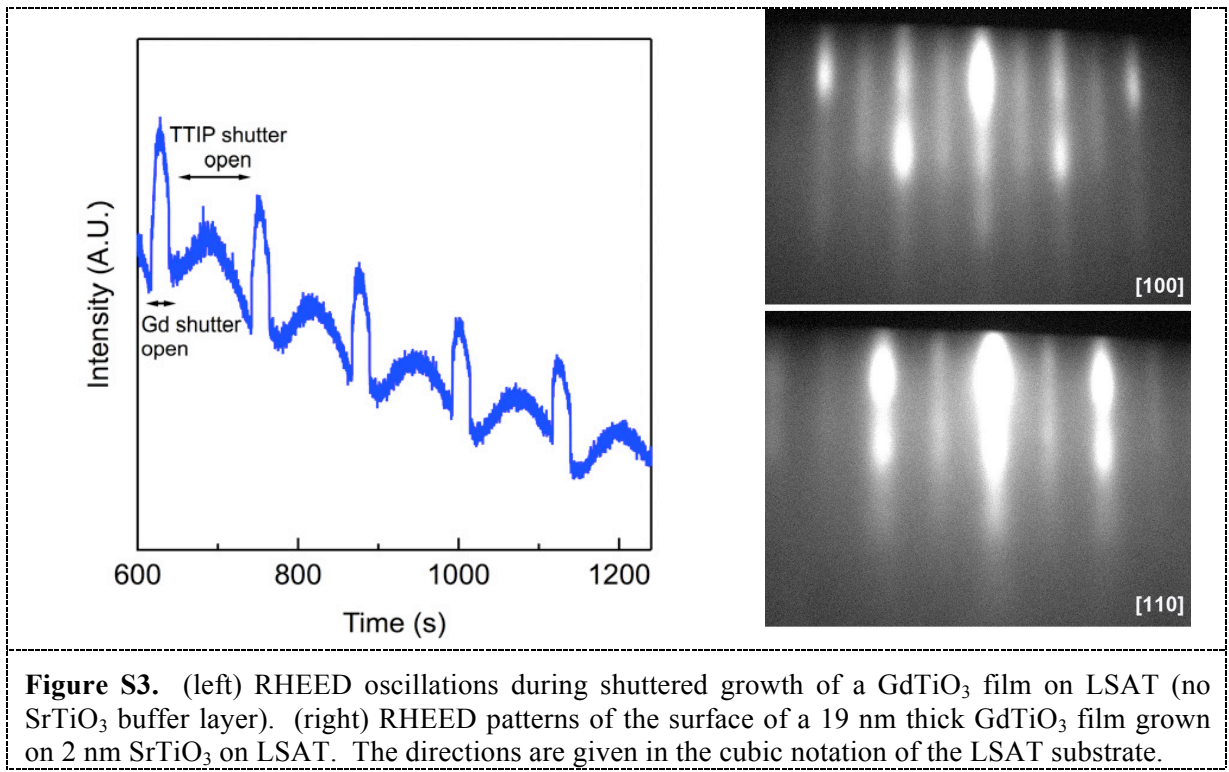

**Figure S3.** (left) RHEED oscillations during shuttered growth of a $GdTiO_3$ film on LSAT (no $SrTiO_3$ buffer layer). (right) RHEED patterns of the surface of a 19 nm thick $GdTiO_3$ film grown on 2 nm $SrTiO_3$ on LSAT. The directions are given in the cubic notation of the LSAT substrate.

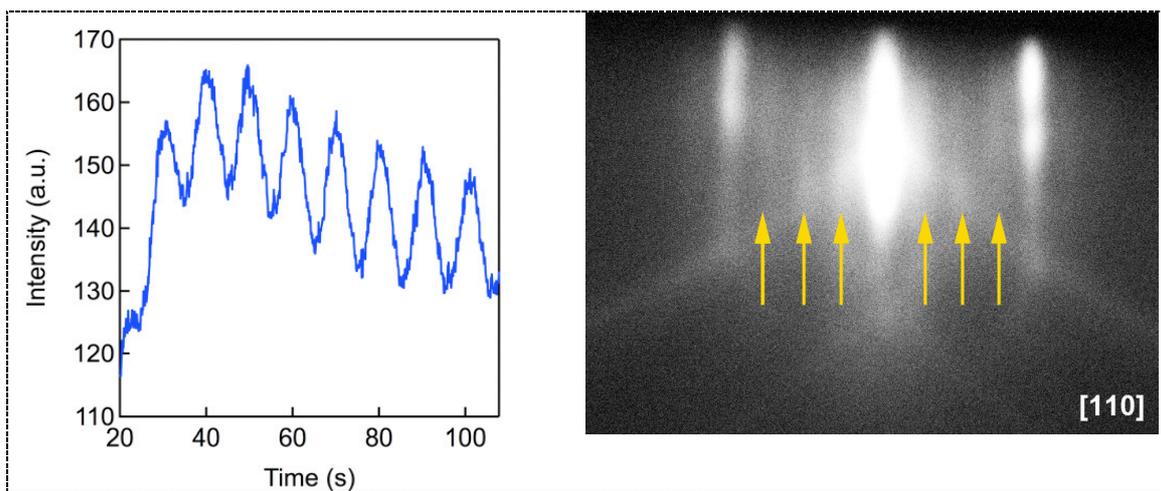

**Figure S4.** (left) RHEED oscillations during growth of SrTiO$_3$ on LSAT. (right) RHEED pattern of the surface of a 60 nm thick SrTiO$_3$ film grown on LSAT.

Figure S5 shows a cross-section high-angle annular dark-field image acquired in aberration-corrected STEM of a thin region of the $x = 20$ multilayer. The thickness of this TEM sample region was about 0.1-0.15 of the total inelastic mean free path.

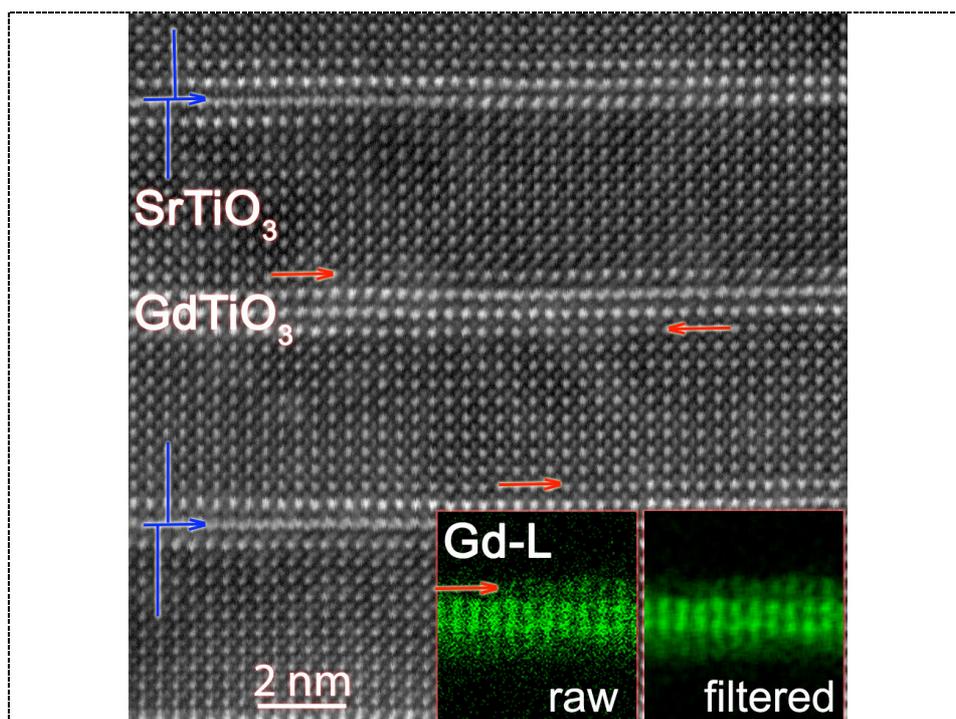

**Figure S5.** Cross-section high-angle annular dark-field image acquired in STEM of a thin region of the $x = 20$ multilayer. The red arrows indicate steps at the interface. The blue lines indicate stacking faults that were occasionally observed in the GdTiO$_3$ layer. The inset shows an atomic resolution energy-dispersive x-ray map of a GdTiO$_3$ layer acquired using the Gd L-edge.

## Hall Resistance

Figure S6 shows as a representative example the Hall resistance of a 19-nm-GdTiO$_3$/44-nm-SrTiO$_3$/LSAT structure as a function of magnetic field at 1.8 K. The Hall resistance as a function of magnetic field is linear. The same linear characteristics were observed for all the other structures and superlattices reported on in this paper.

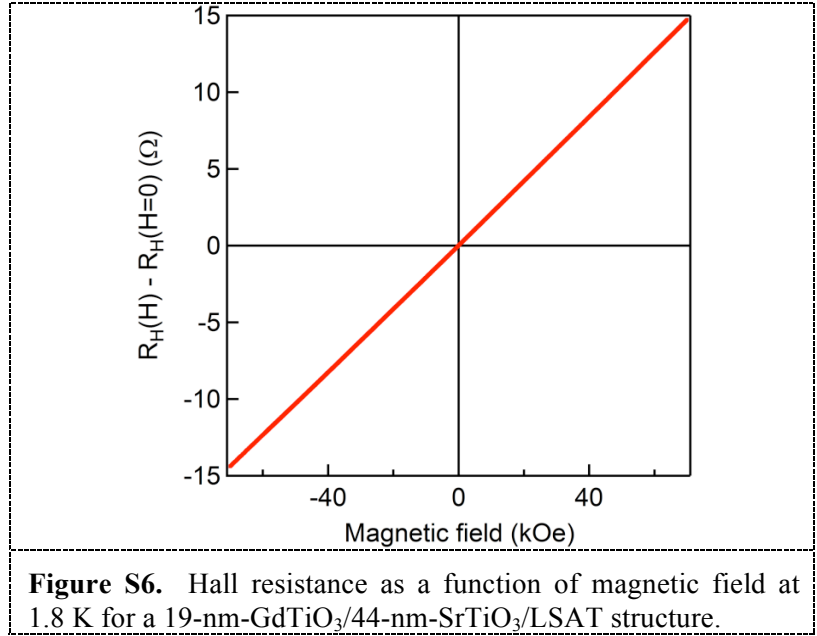

**Figure S6.** Hall resistance as a function of magnetic field at 1.8 K for a 19-nm-GdTiO$_3$/44-nm-SrTiO$_3$/LSAT structure.

## Estimates of the band alignments between SrTiO$_3$ and GdTiO$_3$

SrTiO$_3$ is a band insulator with the states at the valence-band maximum ($E_v$) composed of O $p$ orbitals while those at the conduction-band minimum ($E_c$) are Ti $d$ states. GdTiO$_3$ is a Mott insulator with both $E_v$ and $E_c$ originating from Ti $d$ orbitals. The crystal structure of GdTiO$_3$ was obtained from ref. [2]. The calculations were performed using the hybrid functional method (HSE) [3] as implemented in the VASP code [4]. The bulk of SrTiO$_3$ was simulated using the primitive cell containing 5 atoms with $a_0$ = 3.905 Å, and a 4×4×4 $k$-point mesh for integrations over the Brillouin zone. The bulk of GdTiO$_3$ was simulated using a primitive cell containing 20 atoms to accommodate rotation and tilting of the TiO$_6$ octahedra. A plane-wave basis set of 400 eV was used in the calculations. The SrTiO$_3$/GdTiO$_3$ interface, used to align the averaged electrostatic potential in the two materials, was simulated using a 6×6 superlattice oriented along the [110] direction. The calculated band offsets shown in Fig. S7.

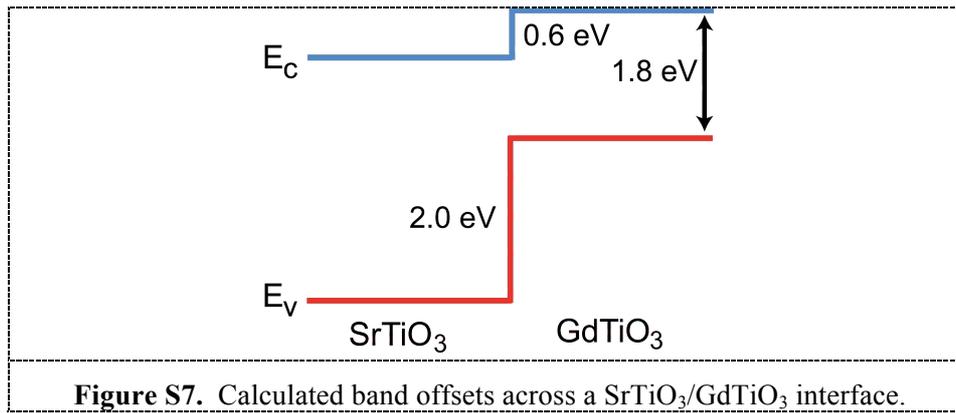

**Figure S7.** Calculated band offsets across a $SrTiO_3/GdTiO_3$ interface.

**References**

[1] B. Jalan, R. Engel-Herbert, N. J. Wright, and S. Stemmer, J. Vac. Sci. Technol. A **27**, 461 (2009).

[2] A. C. Komarek, H. Roth, M. Cwik, W.-D. Stein, J. Baier, M. Kriener, F. Bourée, T. Lorenz, and M. Braden, Phys. Rev. B **75,** 224402 (2007).

[3] J. Heyd, G. E. Scuseria, and M. Ernzerhof, J. Chem. Phys. **118**, 8207 (2003); erratum: J. Chem. Phys. **124**, 219906 (2006).

[4] G. Kresse and J. Furthmüller, Phys. Rev. B **54**, 11169 (1996); G. Kresse and J. Furthmüller, Comput. Mat. Sci. **6**, 15 (1996).